\documentstyle[twocolumn,prl,aps,epsfig]{revtex}
\begin{document}                                                              
\newcommand{\be}{\begin{eqnarray}}
\newcommand{\ee}{\end{eqnarray}}
\newcommand{\dd}{\mathrm{d\,}}
\title{On recent puzzles in the production of heavy quarkonia}
\author{B.L. Ioffe$^{a,b}$ and  D.E. Kharzeev$^b$}
\bigskip
\address{
\bigskip
a) Institute of Theoretical and Experimental Physics,\\
B. Cheremushkinskaya 25, Moscow 117218, Russia\\
\bigskip
b) Department of Physics,\\ Brookhaven National Laboratory,\\
Upton, New York 11973-5000, USA}
\date{June 6, 2003}
\maketitle
\begin{abstract} 

Recently, several surprising experimental observations in the production of heavy quarkonium have been reported.  
In $e^+e^-$ annihilation at $\sqrt{s}=10.6$ GeV, Belle Collaboration finds that $J/\psi$ 
mesons are predominantly produced in association with an extra  
$\bar{c}c$ pair, with $\sigma(e^+e^- \to J/\psi \bar{c}c) / 
\sigma(e^+e^- \to J/\psi X) = 0.59^{+0.15}_{-0.13}\pm 0.12$, and the BaBar collaboration reports that the 
produced $J/\psi$'s have mostly longitudinal  
polarization. In $\bar{p}p$ collisions at the Tevatron, the CDF Collaboration reported an excess of  
$J/\psi$ and $\psi^{\prime}$ mesons at high $p_{\perp}$ over the perturbative QCD predictions; non--perturbative 
approach of NRQCD can accomodate the magnitude of the production cross section but not the observed experimentally 
polarization of quarkonia. 
In this note we propose possible solutions to these puzzles, 
and devise further experimental tests. 

\end{abstract}
\begin{narrowtext}


Experimental and theoretical investigations of heavy quarkonium have been very important in establishing QCD 
as a true theory of strong interactions. The properties of charmonium and bottomonium states have been found in 
qualitative and in some cases quantitative agreement with theoretical expectations based on QCD. 
Nevertheless, since many important non--perturbative problems, including confinement and 
the spontaneous violation of chiral symmetry, 
are still not solved, one may expect that a closer look at the properties and the production of quarkonium 
states will help in elucidating the interplay of short and long distances in this theory. This is why new experimental 
results on quarkonium production always attract much interest and attention.

Recently, Belle Collaboration reported the measurement of $J/\psi$ production in $e^+e^-$ annihilation 
at $\sqrt{s}=10.6$ GeV \cite{Abe:2002rb}. It was found that 
the cross section of $J/\psi$ production significantly exceeds theoretical 
expectations based on the Color Singlet Model (CSM) \cite{CL} and non-relativistic QCD (NRQCD) \cite{SB,YQC}. 
Even more surprisingly, it was found that most of the observed $J/\psi$'s were accompanied by an extra 
$\bar{c}c$ pair, with $\sigma(e^+e^- \to J/\psi \bar{c}c) / 
\sigma(e^+e^- \to J/\psi X) = 0.59^{+0.15}_{-0.13}\pm 0.12$. This ratio exceeds the existing theoretical predictions 
by about an order of magnitude. The problem has recently attracted attention of many theorists -- see, e.g. 
\cite{Brodsky:2003hv,Luchinsky:2003ej,Hagiwara:2003cw,Berezhnoy:2003hz,Liu:2003jj,Bodwin:2002kk,Bodwin:2002fk,AK}.


At first glance, it seems impossible to understand this result within the framework of QCD, where the production of an  
extra $\bar{c}c$ pair should be suppressed both by the large mass of the heavy quarks and by extra powers of the strong 
coupling constant $\alpha_s$. However, this conclusion may be premature. To explain our point, let us consider the 
diagrams of $J/\psi$ production in $e^+e^-$ annihilation. 

There are three types of such diagrams. In the diagrams of the first type (Fig.1a) $J/\psi$  is formed by the fusion 
of $\bar{c}c$ pair produced by the initial virtual photon. This means that $\bar{c}c$ quarks, which initially 
were moving in the opposite directions, turn around and have almost equal and parallel momenta in the final state 
(up to the terms $\sim r_{J/\psi}E_{J/\psi}/M_{J/\psi}$, where $r_{J/\psi}$ and $E_{J/\psi}$ are the radius and the 
energy of $J/\psi$). The momentum conservation is ensured by the radiated gluons which may produce light ($q=u,d,s$) 
$\bar{q}q$ or charmed $\bar{c}c$ pairs. The diagrams of the second type (Fig.1b) correspond to the fragmentation 
of $c$ (or $\bar{c}$) into $J/\psi$. This process requires the production of an additional $\bar{c}c$ pair by 
emission of at  least one gluon by initial $c$ or $\bar{c}$. Finally, the third type of processes are described 
by the diagrams in which the initial virtual photon creates a pair of light quarks and the $J/\psi$ is formed 
from the $\bar{c}c$ pair produced by an exchange of gluons (Fig.1c). Evidently, the diagram of Fig.1c is suppressed 
in comparison to diagrams of Figs.1a,1b by a factor of $\alpha_s(m_c)$ and will be disregarded in what follows. 

On general grounds one may expect that the diagrams of Fig.1b dominate at very high energies of the colliding 
$e^+e^-$, when the ladder of $\bar{c}c$ pairs is formed and the process may be described by Regge theory 
(such process was considered by Kaidalov \cite{AK}). It is easy to estimate the energies starting from which 
one may expect the dominance of the diagrams of Fig.1b. Let $p_c$ be the momentum of the $c-$quark (in the 
$e^+e^-$ c.m. system) fragmenting into $J/\psi$ (see Fig.2). Then the minimal value of the recoil momentum $q$, 
corresponding to the forward production, is equal to
\be
q \simeq {M_{J/\psi}^2 - m_c^2 \over 2 p_c}.
\ee 
By requiring $q$ to be at least as small as $q \simeq 0.5$ GeV (typical for Regge asymptotics), we get 
$p_c > 10 $ GeV, i.e. $\sqrt{s} > 20$ GeV. In fact one may expect that the energy should be $\sqrt{s} > 50$ GeV. 

Fig.1a dominates at low and intermediate $\sqrt{s}$. Our main interest here will be concentrated on the 
qualitative and semi--quantitative 
consideration of its contribution at intermediate $\sqrt{s} \sim 10$ GeV where the measurements of 
Belle and BaBar were done.

\begin{figure}[b]
\epsfig{file=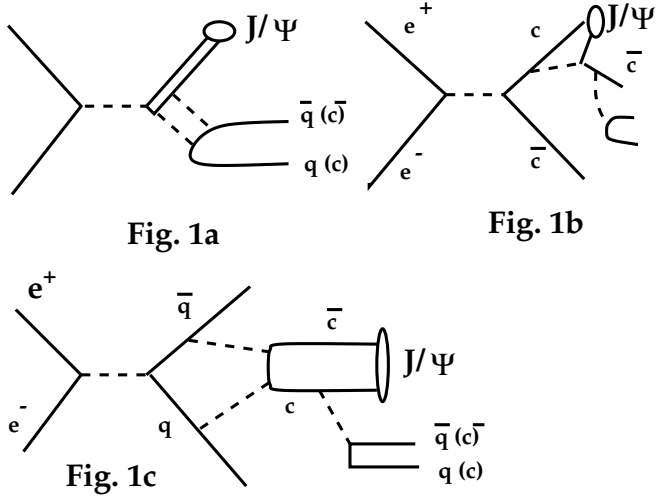, height=2.6in}
\vskip0.1cm
\caption{The production of $J/\psi$ in $e^+e^-$ annihilation.}
\label{psi}
\end{figure}

We will use the well--established notion 
of duality between the quark--antiquark continuum of pQCD and the spectrum of physical bound states, and assume that 
in order to produce the $J/\psi$ (or any bound charmonium state in general) the integral over the invariant mass 
$\sqrt{s_c}$  
of the $\bar{c}c$ pair in the diagrams of Fig.1 should be restricted to small values
corresponding to the typical 
velocity of charm quark in the $J/\psi$ wave function, $v_{\psi}  \simeq 0.45$. Using $v = \sqrt{1-4m_c^2/s_c}$, and 
the pole mass of the charm quark $m_c \simeq 1.3$ GeV, one finds that 
this corresponds to the invariant mass $\sqrt{s_c} = 2 m_c / (1-v_{\psi}^2)^{1/2} \simeq M_{J/\psi}$.

\begin{figure}[b]
\epsfig{file=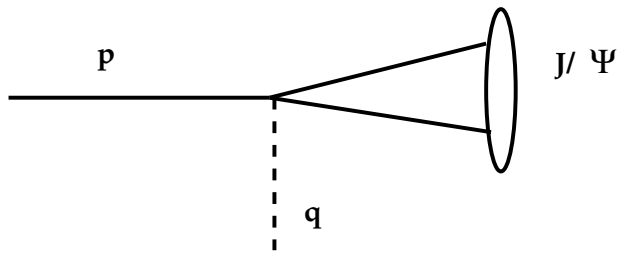, height=1.3in}
\vskip0.1cm
\caption{The kinematics of fragmentation into $J/\psi$.}
\label{frag}
\end{figure}

This means that the invariant mass of the hadron system produced in association with $J/\psi$ in the 
$e^+e^- \to J/\psi X$ process at $\sqrt{s} = 10.6$ GeV can be quite large. 
Belle Collaboration \cite{Abe:2002rb} measured the distribution of events as a function of the mass of the 
system recoiling against the $J/\psi$. The recoil mass was defined as 
\be
M_{recoil}=\sqrt{(\sqrt{s}-E_{J/\psi}^*)^2 - p_{J/\psi}^{*2}}
\ee
where $E_{J/\psi}^*$ and $p_{J/\psi}^*$ are the energy and momentum of $J/\psi$ in the c.m. frame. It was found 
that $M_{recoil}$ concentrates in the domain $M_{recoil} \geq 5$ GeV, with almost no events below $2.8$ GeV 
\cite{Abe:2002rb}. 

We note that this invariant mass is sufficiently larger than the threshold for the production of an additional 
$\bar{c}c$ pair.  
Let us now estimate the relative probabilities of the associated 
with $J/\psi$ production of heavy and light quark--antiquark 
pairs.

As the recoiling invariant mass available for the extra quark pair production is large,   $M_{recoil} \geq 5$ GeV, 
let us, in the first rough approximation, treat $c$ quarks as massless. 
Since three different flavors $(q=u,d,s)$ of light quark pairs 
can be produced, in perturbation theory one should sum over them incoherently, and we expect $\sigma(J/\psi \bar{c}c)/\sigma(J/\psi X) \simeq 1/4$ for $X= \bar{c}c + \bar{q}q$.     
Non-perturbative effects, however, can induce correlations in the production of quark pairs. In this case, 
since the $\bar{q}q$ pairs are produced by gluons, the wave function 
of the light quark state $X_{light}$ has to be an $SU(3)$ singlet:
\be
|X_{light} \rangle = { |\bar{u}u + \bar{d}d + \bar{s}s \rangle \over \sqrt{3}}.
\ee
Therefore, light and charm quarks should be produced in association with the $J/\psi$ with equal probabilities, and we find 
$$ R=\sigma(e^+e^- \to J/\psi \
\bar{c}c) / \sigma(e^+e^- \to J/\psi \ X) = $$ $$ {|\langle e^+e^-
| J/\psi \ \bar{c}c \rangle |^2 \over |\langle e^+e^- | J/\psi \
\bar{c}c \rangle |^2 + {1 \over 3} |\langle e^+e^- | J/\psi \
\bar{u}u + \bar{d}d + \bar{s}s \rangle |^2} \simeq$$
\be
\simeq{1 \over 2} \Biggl [1-O\Biggl
(\frac{4m^2_c}{M^2_{recoil}}\Biggr )\Biggr ], \label{est} \ee
The result (\ref{est}) corresponds to the assumption of maximal coherence in the production of 
light quarks. Under the opposite assumption of completely incoherent production of light 
flavors we would have $R \simeq 1/4$.
Basing on the data \cite{Abe:2002rb} $M_{recoil}$ may be estimated
as $M_{recoil}\approx 5.5$ GeV and $4m^2_c/M^2_{recoil}\approx
0.2$. Therefore, we estimate $m_c$ corrections in (4) as
$O(4m^2_c/M^2_{recoil})\sim 0.2$ and our expectation for $R$ is
\be
R\approx 0.4 \label{qqn}\ee which is not far from the Belle result.

Theoretically the suppression of the spectrum at $M_{recoil} < 3$ GeV may
be understood if we account for the fact that the two gluon state
recoiling against $J/\psi$ is a scalar; because of the scale anomaly in this channel,  
there are strong nonperturbative effects resulting in the high mass of the produced 
gluonic states.

In addition to the diagram of Fig.1a, at sufficiently large $\sqrt{s}$ one should also consider the diagram of charm 
quark fragmentation, see Fig.1b. This diagram always produces an extra $\bar{c}c$ pair in association with the $J/\psi$, 
giving the ratio (\ref{est}) equal to unity, and the admixture of the heavy quark fragmentation will thus increase our 
estimate for the ratio (\ref{est}).   

Let us now turn to the discussion of $J/\psi$ polarization. Since the photon in $e^+e^-$ annihilation is mostly 
transverse, it has helicity $\lambda = \pm 1$. Therefore the $\bar{c}c$ produced by the photon should have 
opposite helicities of $\lambda_c= + 1/2, \lambda_{\bar{c}} = -1/2$ or $\lambda_c= - 1/2, 
\lambda_{\bar{c}} = + 1/2$. 
Initially, 
heavy quark and antiquark move in the opposite directions in the center of mass system of $e^+e^-$ annihilation 
with the velocities $v = \sqrt{1-4m_c^2/s}$ which are close to $v \simeq 1$ at $\sqrt{s} = 10.6$ GeV. 
However, to become bound in 
the $J/\psi$ (or any other bound state of charmonium), at least one of the quarks has to change the direction 
of its momentum by radiating gluons (and extra quark--antiquark pair(s)) since the relative velocity of heavy quarks in a 
bound state should be small. Since the coupling of quarks to gluons $\bar{q}\gamma_{\mu}q A_{\mu}$ in QCD 
conserves the helicity of the quark, a change in the direction of its momentum should be acoompanied by the spin flip. 
We thus come to the conclusion that in the case of  $J/\psi$ produced at high momentum, 
the total spin of $J/\psi$ should have zero projection on its direction of 
motion, which corresponds to the longitudinal polarization. 

This is in agreement with the experimental result of the BaBar Collaboration, which states that the angular 
distribution of positively charged lepton decay product with respect to the direction of $J/\psi$ measured in the CM 
frame is $W(\theta) \sim 1 + \alpha \ cos^2 \theta$ with $\alpha = -0.46 \pm 0.21$ for CM momentum $p^* < 3.5$ GeV, 
and $\alpha = -0.80 \pm 0.09$ for CM momentum $p^* > 3.5$ GeV. (In this distribution, 
$\alpha = -1$ corresponds to longitudinal polarization, $\alpha = +1$ to transverse, and $\alpha = 0$ indicates no 
polarization).

Let us discuss now the case of very high $\sqrt{s} > 50$ GeV, when the dominant process of $J/\psi$ production 
is the $c (\bar{c})$ quark fragmentation, see Fig.1b. It is evident that in this case any produced $J/\psi$ 
is acompanied by an extra $\bar{c}c$ pair, and therefore the ratio
$$
\sigma(e^+e^- \to J/\psi \bar{c}c) / 
\sigma(e^+e^- \to J/\psi X) \simeq 1.
$$
As is clear from Fig.1b one of the $c$ quarks has helicity, say, $\lambda = +1/2$; 
the helicity of the other quark created from the vacuum is uniformly distributed. Therefore the mean value of 
$\alpha$ is equal to zero (it is easy to see that the cases of $\alpha=+1$ and $\alpha=-1$ have equal 
probabilities), and the produced $J/\psi$ is unpolarized. 

These two observations show that the fragmentation mechanism cannot describe the data \cite{Abe:2002rb}, 
 \cite{Aubert:2001pd} at $\sqrt{s}=10.6$ GeV, in agreement with our expectations.

A similar mechanism can be expected to work in $J/\psi$ and $\psi'$ production at high $p_{\perp}$. 
Indeed, the dominant mechanism of $\bar{c}c$ production in this kinematical region is the gluon fragmentation 
\cite{Braaten:1993rw} ; 
at high $p_{\perp} \gg 2 m_c$ we may assume that the gluon is almost on shell and thus is transversely polarized. 
Apart from the overall color charge, the process thus becomes analogous to the $e^+e^-$ annihilation that we 
considered above, and we expect that the same conclusion about the longitudinal polarization of $J/\psi$ and $\psi'$ 
should hold there as well. We thus predict that the polarization of $J/\psi$ and $\psi'$, as well as of bottomonium 
states, will become increasingly longitudinal with the increase of $p_{\perp}$.
This observation seems to be consistent with the CDF data \cite{Affolder:2000nn,Acosta:2001gv}, 
but better statistics and a larger range in $p_{\perp}$ 
would be needed to establish this conclusively. 
Recently it was suggested that the fragmentation of charm quarks can dominate the production of $J/\psi$ and $\psi'$ 
at $p_{\perp}$ \cite{Qiao:2002nd}; if this is the case, then as we argued earlier, 
no polarization should be present 
at high $p_t$.

B.I. is thankful to Larry McLerran for the hospitality at Brookhaven National Laboratory, where this 
work was done. The work of B.I. was supported in part by INTAS Grant 2000--587 and RFBR Grant 03--02--16209. 
D.K. is grateful to Maurice Goldhaber for stimulating discussions. 
The research of D.K. is supported by the U.S. Department of Energy under Contract No. DE-AC02-98CH10886.

\end{narrowtext}
\end{document}